\documentclass[aps,preprintnumbers,prl,twocolumn,superscriptaddress]{revtex4}

\usepackage{slashed}
\usepackage{graphicx,color}
\usepackage{epsfig}
\usepackage{subfigure}
\usepackage{epsfig}
\usepackage{dcolumn}
\usepackage{bm}
\usepackage{color}

\begin{document}

\preprint{ACFI-T16-03}

\title{The Diphoton Excess Inspired  Electroweak Baryogenesis}

\author{Wei Chao}
\email{chao@physics.umass.edu}

 \affiliation{ Amherst Center for Fundamental Interactions, Department of Physics, University of Massachusetts-Amherst
Amherst, MA 01003, United States }

\vspace{3cm}

\begin{abstract}

A resonance in the diphoton channel with invariant mass $m_{\gamma \gamma}^{} = 750 ~ {\rm GeV}$ was claimed by the ATLAS and CMS collaborations at the run-2 LHC with $\sqrt{s}=13~{\rm TeV}$.  
In this paper, we explain this diphoton excess as a pseudo-scalar singlet, which is produced at the LHC through gluon fusion with exotic scalar quarks running in the loop.
We point out the scalar singlet  might trigger a two-step electroweak phase transition, which can be strongly first oder. 
By assuming there are CP violations associated with interactions of the scalar quarks, the model is found to be able to generate adequate baryon asymmetry of the Universe through the electroweak baryogenesis mechanism.  
Constraints  on the model are studied.

\end{abstract}

\maketitle
\noindent {\bf {\color{blue}Introduction}} 
An excess in the diphoton channel with invariant mass $m_{\gamma \gamma} \approx 750~{\rm GeV}$ was observed by both the ATLAS~\cite{ATLAS13} and CMS~\cite{CMS13} collaborations at the run-2 LHC.  The significance is $3.6\sigma$ for the ATLAS and $2.6\sigma$ for the CMS. 
According to the Landau-Yang theorem~\cite{Landau:1948kw,Yang:1950rg}, the resonance might be a spin-0/2 state.
The gluon fusion or heavy quark production of the resonance is favored since the run-1 LHC did not see the resonance.
There are many interesting interpretations~~\cite{Angelescu:2015uiz,Backovic:2015fnp,Bellazzini:2015nxw,Buttazzo:2015txu,DiChiara:2015vdm,Ellis:2015oso,Franceschini:2015kwy,Gupta:2015zzs,Harigaya:2015ezk,Higaki:2015jag,Knapen:2015dap,Low:2015qep,Mambrini:2015wyu,McDermott:2015sck,Molinaro:2015cwg,Nakai:2015ptz,Petersson:2015mkr,Pilaftsis:2015ycr,Dutta:2015wqh,Cao:2015pto,Matsuzaki:2015che,Kobakhidze:2015ldh,Martinez:2015kmn,Cox:2015ckc,Becirevic:2015fmu,No:2015bsn,Demidov:2015zqn,Chao:2015ttq,Fichet:2015vvy,Curtin:2015jcv,Bian:2015kjt,Chakrabortty:2015hff,Agrawal:2015dbf,Csaki:2015vek,Falkowski:2015swt,Aloni:2015mxa,Bai:2015nbs,Chao:2015nsm,Han:2015cty,xxx,Ibanez:2015uok,Arun:2015ubr,Chao:2016mtn,Ito:2016zkz} of the excess as manifestations of various new physics models.
The diphoton excess, should it exist, would offer new possibilities for addressing problems that can not be solely solved in the Standard Model (SM) of particle physics.

The origin of the baryon asymmetry (BAU) is one unsolved problem in cosmology. 
From Planck~\cite{Ade:2015xua} and WMAP~\cite{Komatsu:2010fb}, one has $Y_B \equiv \rho_B /s =(8.59\pm0.11)\times 10^{-11}$, where $\rho_B$ and $s$ are the baryon number density and entropy density respectively.   
Starting with a matter-antimatter symmetric Universe at the end of the inflationary epoch, new dynamics  is needed to generate the observed BAU during the subsequent cosmological evolution, which should satisfy three Sakharov criteria~\cite{Sakharov:1967dj}: (1) baryon number violation; (2) C and CP violation; (3) A departure from the thermal equilibrium.  
Among all available  scenarios, electroweak baryogenesis mechanism (EWBG)~\cite{Morrissey:2012db} is the most attractive one, as it might be tested at both high energy colliders and low energy electric dipole moment (EDM) measurements~\cite{Engel:2013lsa}.

In this letter, we explain both the $750$ GeV diphoton excess and the baryon asymmetry in a simple model as a prototype of new physics behind.  
The resonance is explained as the CP-odd component of  a scalar singlet $S$, produced via loop induced gluon fusion with scalar quarks running in the loop. 
$S$ on the other hand might trigger a  two-step electroweak phase transition (EWPT) as required by the EWBG. 
CP-violating interactions of scalar quarks on the expanding bubble wall may generate nonzero charge densities, diffused into the plasma, which can be converted into a density of SM Higgs via the trilinear interaction, and then be translated into  a net densities of left-handed fermions $n_L$ via inelastic Yukawa interactions.   
While $n_L$ serves as the source of generating baryon asymmetry through the weak sphaleron process. 
As will be seen in the next section the model might be embedded into a general NMSSM model~\cite{Ellwanger:2009dp} extended with vector-like quarks.  
Constraints on the model are briefly discussed.

\noindent {\bf {\color{blue}The model}} 
We extend the SM with a complex scalar singlet $S$, an exotic scalar quark doublet $\tilde{Q}$ transforming as $(3, ~2,~7/6)$ under the SM gauge group $SU(3)_C\times SU(2)_L \times U(1)_Y$, and a scalar quark singlet $\tilde T$, transforming as $(3,~1,~2/3)$. $\tilde Q$ and $\tilde T$ can superpartners of the certain vector-like femrions.
The Higgs potential can be written as 
\begin{eqnarray}
V&=& -\mu_h^{2} H^\dagger H + \mu_s^2 S^\dagger S +\lambda (H^\dagger H)^2 + \lambda_1 S^\dagger S H^\dagger H 
\nonumber \\ &&
+ \lambda_s^{} (S^\dagger S)^2 +\{ -\mu_a^2  S^2 + \lambda_2  S^2  H^\dagger H + h.c.\} \; ,  \label{potential}
\end{eqnarray}
where $H=(H^+,~\{h+v+i G\}/\sqrt{2})^T$ and $S= ( s+i a )/\sqrt{2}$, with $v=246~{\rm GeV}$ from precision measurements.
We further assume there is no CP violation in the potential, such that $\mu_a^2$ and $\lambda_2$ are both real. 
Interactions between the Higgs and scalar quarks take the following form
\begin{eqnarray}
{\cal L} &\ni&  A_1^{} S \tilde T^\dagger \tilde T + A_2^{} S \tilde Q^\dagger \tilde Q + A_3^{} \tilde Q^\dagger H \tilde T  +\nonumber \\ && M_{\tilde T}^2 \tilde T^\dagger \tilde T +M_{\tilde Q}^2 \tilde Q^\dagger \tilde Q + a_4^{} \tilde Q^\dagger HS\tilde T+ {\rm h.c.} \; .\label{matterint}
\end{eqnarray} 
Interactions in Eq. (\ref{matterint}) can actually embedded into a general NMSSM~\cite{Ellwanger:2009dp} extended with vector like fermions $\hat Q,~\hat T$, where new terms in superpotential are
\begin{eqnarray}
W&\ni& \hat\lambda \hat S \hat H_u \hat H_d + Y \hat { Q} \hat H_d \hat T_R^c + Y_2 \hat S \hat T_L \hat T_R^c + Y_3 \hat S \hat Q_L^{} \hat Q_R^c \nonumber \\
&& + M_1 \hat T_L \hat T_R^c + M_2 \hat Q_L \hat Q_R^c \; . \label{superpotential}
\end{eqnarray}
In this case terms proportional to $A_{1,2,3,4}$ in Eq. (\ref{matterint}) come from $F$-terms, while mass terms come from the soft supersymmetry breaking Lagrangian.  
There are some other $F$-terms arsing from Eq. (\ref{superpotential}), we omit them for simplification.   
In the following, we only focus on the phenomenology induced by Eqs. (\ref{potential}) and (\ref{matterint}), leaving a systematic study of the full theory to a future work.
%
%

By assuming $\mu_s^2 -2 \mu_a^2 + {1\over 2 } (\lambda_1+2\lambda_2) v^2 >0$, $S$ gets no vacuum expectation value (VEV) at the zero temperature. 
The mass eigenvalues of $h$, $s$ and $a$ can be written as
$
m_h^2 = 2 \lambda v^2 \; , ~ m_s^2 = \mu_s^2 -2 \mu_a^2 + {1\over 2 } (\lambda_1+2\lambda_2) v^2 \; ,  
~ m_a^2 = \mu_s^2 +2 \mu_a^2 + {1\over 2 } (\lambda_1-2\lambda_2) v^2 \; 
$.
Apparently the mass difference of $a$ and $s$ can be arbitrary. 
In the rest of this paper we set $\lambda_2 = 0 $ for simplification. 
There is no mixing between scalars. 


\begin{figure}[h]
  \includegraphics[width=0.3\textwidth]{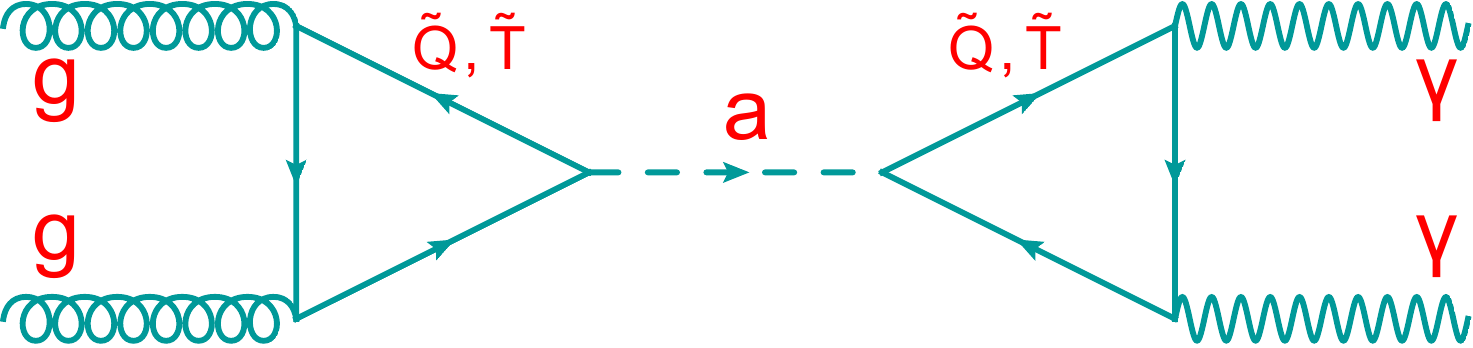}
\caption{\label{feyn}  Feynman diagram for the production of the resonance at the LHC.}
\end{figure}

\noindent {\bf {\color{blue}Diphoton excess}} 
Both the ATLAS and CMS collaborations have observed the diphoton excess of invariant mass $m_{\gamma \gamma } \approx 750 ~{\rm GeV}$ at the run-2 LHC with center-of-mass energy $\sqrt{s}=13 ~{\rm TeV}$.
The cross sections of the excess can be estimated as $\sigma(pp\to \gamma \gamma) \approx $ $(6\pm 3) ~{\rm fb}$ for CMS and $(10\pm3)~ {\rm fb}$ for ATLAS.
In this section we explain this resonance as the signal of the CP-odd scalar $a$ in our model. 
It can be produced via gluon fusion and decay into diphoton with scalar quarks running in the loop.
The relevant feynman diagram is given in Fig.~\ref{feyn}.  
The signal cross section at the LHC can be written as
\begin{eqnarray}
\sigma(pp\to  a\to\gamma \gamma ) ={1\over sM_a \Gamma_a } C_{gg} \Gamma_{gg}^{} {\Gamma}_{\gamma \gamma}^{} 
\end{eqnarray}
where $M_a$ and $\Gamma_a$ are the mass eigenvalue and total rate of $a$ respectively; $C_{gg}$ is the dimensionless partonic integral and one has $C_{gg}\approx 3163$~\cite{Franceschini:2015kwy} at $\sqrt{s}=13~{\rm TeV}$;  $\Gamma_{gg}$ and $\Gamma_{\gamma \gamma}$ are decay rates of $a$ to digluon and diphoton respectively, which take the following form 
\begin{eqnarray}
\Gamma_{gg} &=& {\alpha_s^2 M^3 \over 128 \pi^3 } \left| \sum_{i=1}^2 {\sqrt{2} (1+\delta_{i2}^{} ) {\rm Im} A_i \over  2  M_i^2 } f(\tau_i^{} )   \right|^2  \; ,\label{rategg}\\
\Gamma_{\gamma \gamma } &=&{\alpha_e^2 M^3 \over 1024 \pi^3 } \left| \sum_{i=1}^2 {\sqrt{2} C_i^{} {\rm Im} A_i \over    M_i^2 } f(\tau_i^{} )   \right|^2   \; ,    \label{ratepp}
\end{eqnarray}
with $C_1=4/3$ and $C_2 = 29/3$. The expression of $f(x)$ can be found in Ref.~\cite{Gunion:1989we}. $s$ might also decay into diphoton or digluon with the amplitude proportional to ${\rm Re} A_i$.  Since LHC did not see any other diphoton excess, we assume $A_i$ is purely imaginary.

\begin{figure}[t]
  \includegraphics[width=0.45\textwidth]{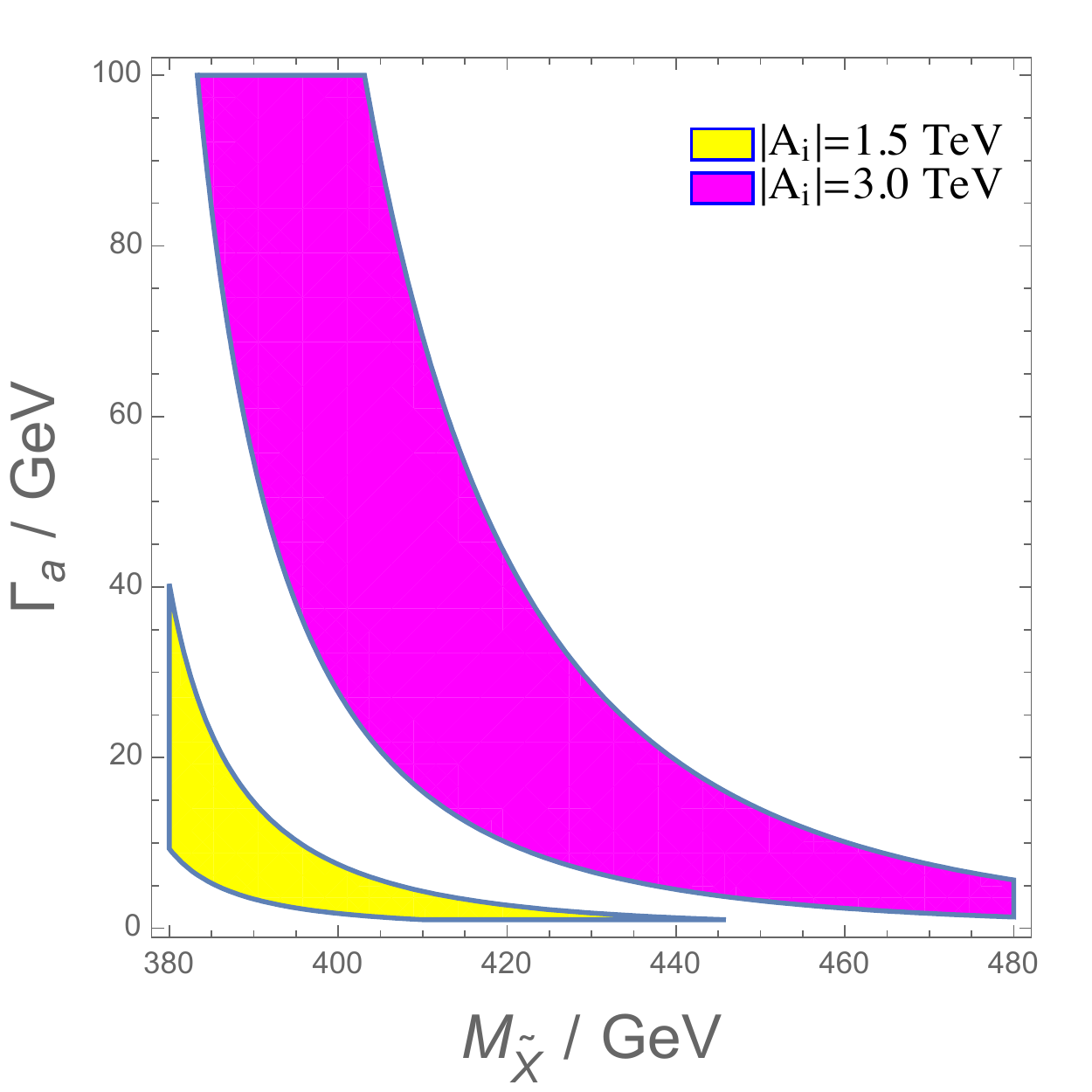}
\caption{\label{signal}  Regions in the $M_{\tilde X} -\Gamma_a$ plane that have $\sigma(gg\to a \to \gamma \gamma) \in (3,~13)~{\rm fb}$ for $|A_i|=1.5~{\rm TeV}$ (yellow) and $3.0~ {\rm TeV}$ (magenta) respectively.}
\end{figure}

From Eqs. (\ref{rategg},\ref{ratepp}), one might estimate the size of the ratio $r\equiv \Gamma_{gg}/\Gamma_{rr}$, which is the function of $|A_i|$ and $M_{\tilde T, \tilde Q}$. 
By assuming $ |A_1|\approx |A_2|$ and $M_{\tilde T } \approx M_{\tilde Q}$ one has $r=19.6$.  
Using the same assumptions, we plot in the $M_{\tilde T} -\Gamma_a$ plane the region that has $\sigma(gg\to a \to \gamma \gamma ) \in (3,~13) ~{\rm fb}$. The magenta region corresponds to $|A_i|=3.0~{\rm TeV}$ while the yellow region corresponds to $|A_i| =1.5~{\rm TeV}$.  
It is clear that the best fit width,  $\Gamma =45~{\rm GeV}$ from ATLAS, can be realized in this model only for relatively large $|A_i|$ scenario. 
So far $a$ can only decay to digluon and diphoton. 
Extra decay channels might open if there is CP violation in the potential. 
For example, $\alpha |S^2| S + \beta S H^\dagger H$ terms may lead to the process $a\to 2s, 2h$. 
We leave the study of decays of $a$ to a  future project. 

Finally let us check constraints on the model from the run-1 LHC. 
For a resonance with $M=750~{GeV}$ and $\Gamma=45~{\rm GeV}$, the upper bounds on cross sections of various final states  at the 95\% CL are $\sigma\cdot {\rm BR}_{\rm ZZ}<12~{\rm fb}$~\cite{Aad:2015kna}, $\sigma\cdot {\rm BR}_{\rm WW}<40~{\rm fb}$~\cite{Aad:2015agg}, $\sigma\cdot {\rm BR}_{\rm hh}<39~{\rm fb}$~\cite{Aad:2015xja} and $\sigma\cdot {\rm BR}_{ jj}<2.5~{\rm pb}$~\cite{Aad:2014aqa}.  
The first three bounds constrain the mixing between $a$ and $h$, which is exactly zero for our case.  
For the dijet constraint, the bound is $\Gamma (a\to gg)/\Gamma(a\to \gamma \gamma ) <1222$~\cite{Franceschini:2015kwy}. 
While in our case, $r\sim {\cal O } (20)$, which is far below the current constraint.

\begin{figure}[t]
  \includegraphics[width=0.45\textwidth]{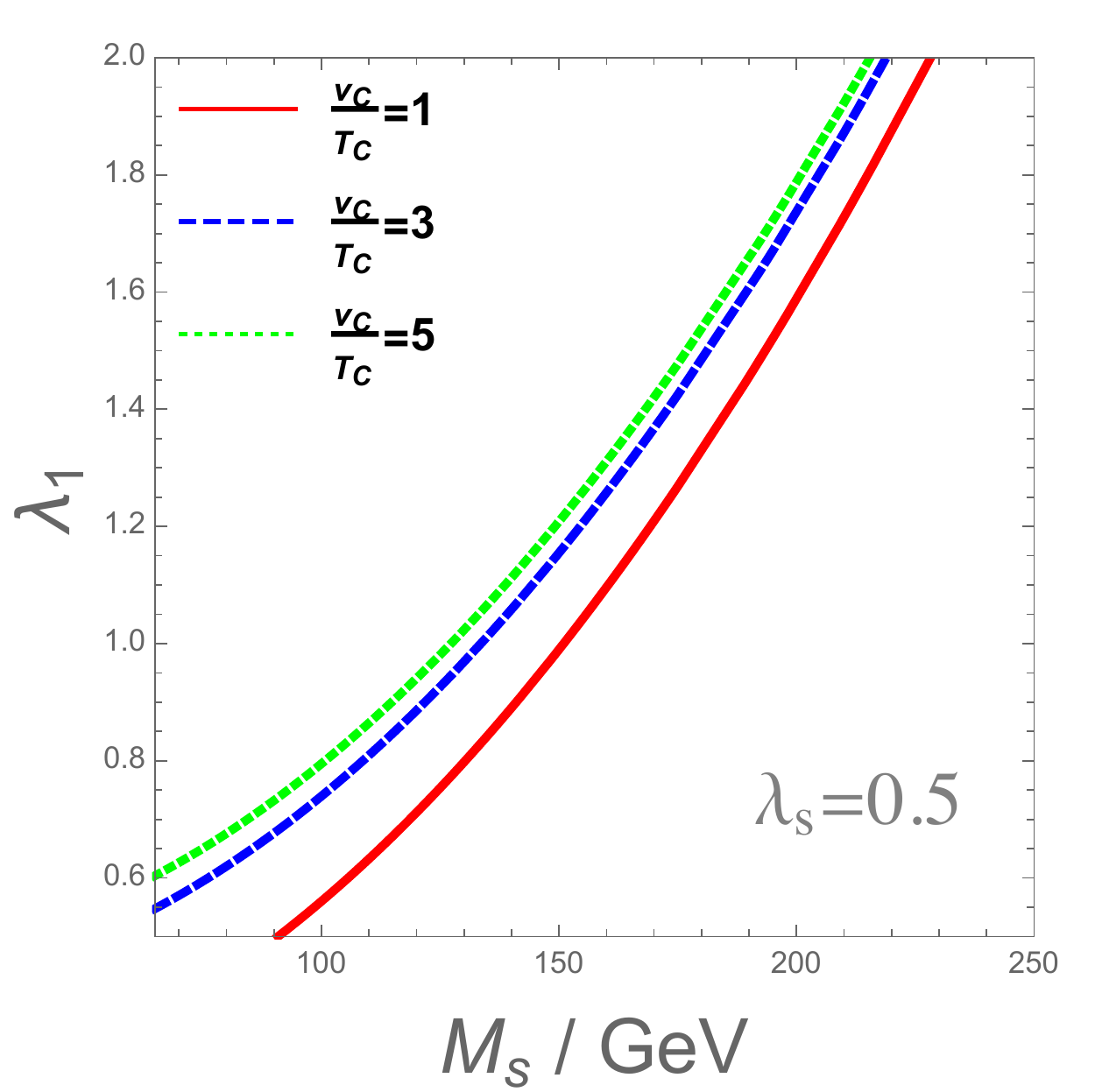}
\caption{\label{phasetrans}  Parameter space in the $m_s-\lambda_1$ plane for the strongly first order EWPT, The solid, dashed and dotted lines correspond to $v_c/T_C=1,3,5$ respectively. }
\end{figure}

\noindent {\bf {\color{blue}EWPT}} 
The dynamics of the EWPT, which occurred about $10^{-10}$ second after the Big Bang, are important in explaining the BAU via the EWBG mechanism.
The condition that the baryon excess generated at the EWPT not be washed out requires a strongly first order EWPT. 
In this model, EWPT can be two step~\cite{Espinosa:2011ax}, where there is large barrier at the tree level between $\langle H\rangle \neq0$, $\langle S \rangle =0 $ and $ \langle H  \rangle =0$, $\langle S\rangle \neq 0$. 
The effective potential at the finite temperature can be written in terms of the background fields $h$ and $s$,
\begin{eqnarray}
V_{T} &\approx&  - {1\over 2 }(\mu_h^2 - \Pi_h) h^2  + {1\over 4 }\lambda h^2  + {1\over 4 } \lambda_1 h^2 s^2  \nonumber \\ &&-{1\over 2 }(2\mu_a^2 -\mu_s^2 -\Pi_s) s^2 + {1 \over 4 } \lambda_s s^4 \; .
\end{eqnarray}
where $\Pi_h$ and $\Pi_s$ are thermal masses of $h$ and $s$ respectively, and can be written as
\begin{eqnarray}
\Pi_h &=& \left\{ {3g^2 + g^{\prime 2 } \over 16} + {\lambda \over 2 } + {y_t^2 \over 4 } + {\lambda_1 \over 12} \right\} T^2   \; , 
\nonumber \\
\Pi_s &=& \left\{ {\lambda_s \over 3 }+ {\lambda_1 \over 6 }\right\} T^2 \nonumber  \; .
\end{eqnarray}
Interactions in Eq. (\ref{matterint}) have negligible contribution to thermal masses for two reasons: (1) only quadratically divergent self-energy diagrams at zero temperature contribute to Debye mass~\cite{Comelli:1996vm}; (2) $M_{\tilde Q, \tilde T}$ are larger than the critical temperature, which is about ${\cal O} (100)~{\rm GeV}$. 

We focus on the parameter space of $2m_a^2-m_s^2 >0$ and ${1/ 2 }\lambda_1 v^2 -2m_a^2 +m_s^2 >0$, where $s$ gets no VEV at $T=0$ but gets non-zero VEV at finite temperature.  
There are two types of minima: $(v_h,~ 0)$ and $(0, ~v_s)$. Conditions for the strongly first order EWPT are
\begin{eqnarray}
V_{T} (v_h(T_C), 0, T_C) = V_{T} (0, v_s(T_C), T_C), \;  \&  \;  {v_h(T_C)\over T_C} \geq1 \; .
\end{eqnarray}

In Fig.~\ref{phasetrans}, we show  in the $m_s-\lambda_1$ plane, the parameter space, that might give rise to a strongly first oder EWPT by assuming $\lambda_s=0.5$. 
The solid, dashed and dotted lines correspond to $v_C/T_C=1,~3,~5$ respectively.
 Region above the solid line is able to generate a strongly first oder EWPT.

\noindent {\bf {\color{blue}EWBG}} 
In this model, three Sarkharov conditions are realized as follows:
First, the two-step EWPT can  be strongly first oder, which guarantees a departure from thermal equilibrium at $T\sim 100~{\rm GeV}$. 
Second, CP-asymmetric charge densities might be produced by the CP-violating interactions of scalar quarks at the expanding bubble wall, where  VEVs are space-time dependent. 
CP asymmetries, diffused into the plasma, can be converted into a net density of left-handed fermions, $n_L$, through inelastic tri-scalar interactions and Yukawa interactions. 
Third, baryon asymmetry will be generated through the weak sphaleron process in the presence of nonzero $n_L$, which, once captured by the expanding bubbles, will be preserved since strongly first oder EWPT  quenches the sphaleron transitions inside the bubble.

\begin{figure}[t]
  \includegraphics[width=0.15\textwidth]{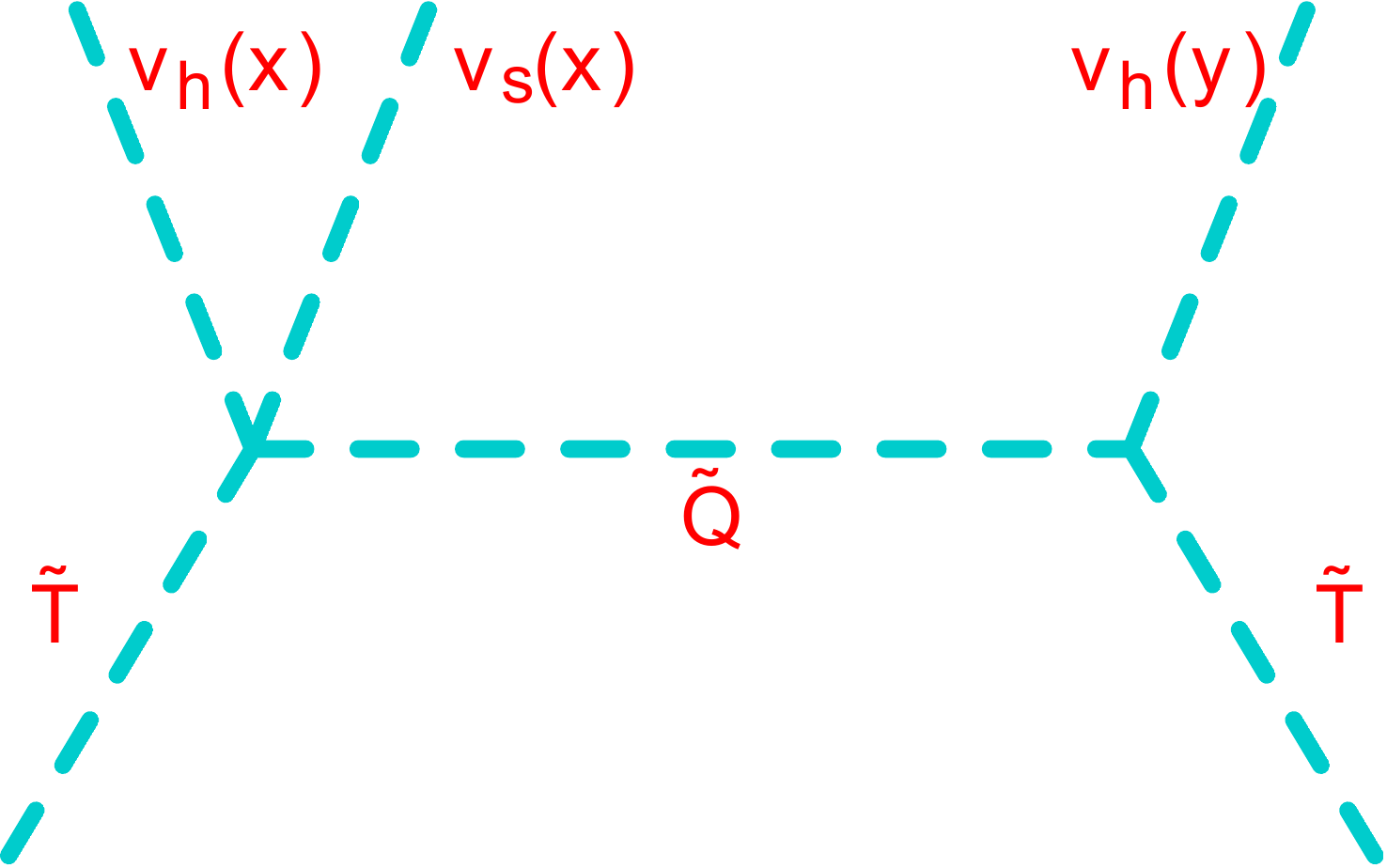}
  \hspace{0.3cm}
  \hspace{0.3cm}
  \includegraphics[width=0.15\textwidth]{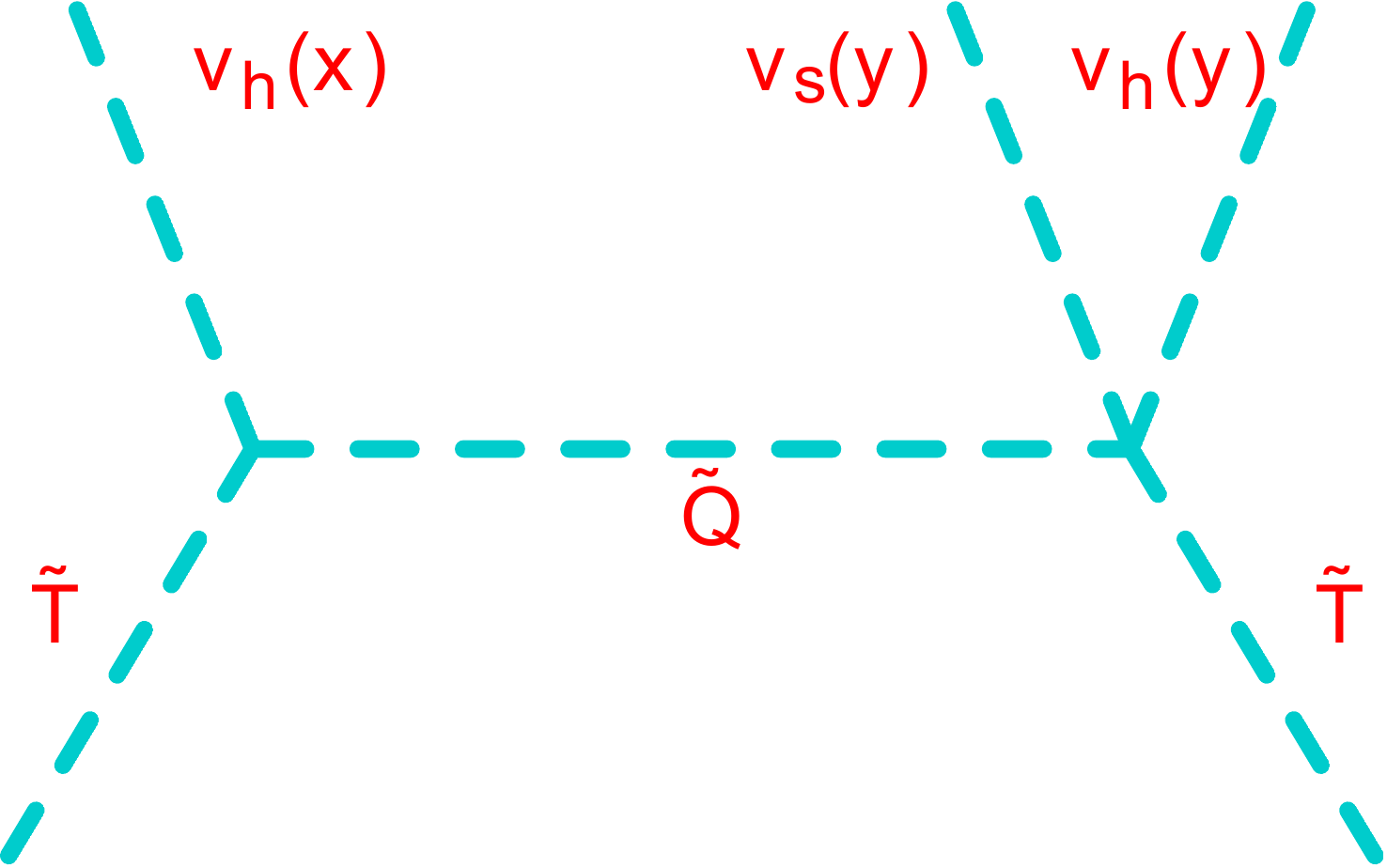}
\caption{\label{baryogenesis}  Diagrams for the CP-violating source terms.}
\end{figure}

Working in the closed-time-path formalism~\cite{Chou:1984es}, we calculate  CP-violating source terms under the ``VEV-insertion" approximation~\cite{Lee:2004we,Riotto:1998zb}.
Relevant diagrams are given in Fig.~\ref{baryogenesis}. 
The source term can then be written as
\begin{eqnarray}
S^{\rm CPV}_{\tilde T } &=& {N_c   v^2 \dot{v}_s \over 2\pi^2 } {\rm Im} [A_3^* A_4^{}]\int_0^{\infty} {k^2 d k \over \omega_{\tilde  Q} \omega_{\tilde T}} \times \\
&&{\rm Im} \left\{ { n_B(\varepsilon_{\tilde T}^*) -n_B(\varepsilon_{\tilde Q}^{} ) \over (\varepsilon_{\tilde Q}^{}-\varepsilon_{\tilde T}^*  )^2} + { n_B(\varepsilon_{\tilde T}^{}) +n_B(\varepsilon_{\tilde Q}^{} ) \over (\varepsilon_{\tilde Q}^{}+\varepsilon_{\tilde T}^{} )^2} \right\} \nonumber 
\end{eqnarray}
where $N_c=3$ being the color factor, $\omega_X^2 = m_X^2 + k^2$ with $m_X$ the thermal mass of $X$,
$\varepsilon_X=\omega_X - i\Gamma_X$ with $\Gamma_X$ the decay rate of $X$. 
Transport equations can be written as
\begin{eqnarray}
\partial_\mu ^{} \tilde T^\mu &=&+ {\Gamma_{\tilde Q}^\pm  } \left(  {\tilde Q  \over k_{\tilde Q} } \pm {{\tilde T \over k_{\tilde T}}}\right) -\Gamma_{tri }^{H} \left({{\tilde T \over k_{\tilde T} } + {H \over k_H } - { \tilde Q \over k_{\tilde Q}} } \right) \nonumber \\
&&- \Gamma_{tri}^{S\tilde T} {S \over k_S }+S_{\tilde T}^{\rm CPV} \label{diff1}  \\
\partial_\mu ^{} \tilde Q^\mu &=& -{\Gamma_{\tilde Q}^\pm  } \left(  {\tilde Q  \over k_{\tilde Q} } \pm {{\tilde T \over k_{\tilde T}}}\right) +\Gamma_{tri }^{H} \left({{\tilde T \over k_{\tilde T} } + {H \over k_H } - { \tilde Q \over k_{\tilde Q}} } \right) \nonumber \\
&&- \Gamma_{tri}^{S\tilde Q} {S \over k_S }-S_{\tilde T}^{\rm CPV}\label{diff2}    \\
\partial_\mu^{} H^\mu &=& -\Gamma_h^{} {H \over k_H } - \Gamma_Y^t \left( {Q \over k_Q} + {H \over k_H } - {T \over k_T } \right) \nonumber \\
&& -\Gamma_{tri }^{H} \left({{\tilde T \over k_{\tilde T} } + {H \over k_H } - { \tilde Q \over k_{\tilde Q}} } \right)  \label{diff13}  \label{diff4}\\
\partial_\mu^{} S^\mu &=& - \Gamma_{tri}^{S\tilde Q} {S \over k_S }- \Gamma_{tri}^{S\tilde T} {S \over k_S }  \label{diff5}  \\
\partial_\mu T^\mu &=&-\Gamma_m^t \left(  {T \over k_T } -{Q \over k_Q }\right) -\Gamma_Y^t \left( {T \over k_T } -{Q\over k_Q } -{H \over k_H } \right) \nonumber \\
&& -\Gamma_{\rm ss}^{} \left( {T \over k_T } - 2 {Q \over k_Q } + 9 {B \over k_B} \right) \label{diff6}   \\
\partial_\mu Q^\mu &=&+\Gamma_m^t \left(  {T \over k_T } -{Q \over k_Q }\right) +\Gamma_Y^t \left( {T \over k_T } -{Q\over k_Q } -{H \over k_H } \right) \nonumber \\
&& +2\Gamma_{\rm ss}^{} \left( {T \over k_T } - 2 {Q \over k_Q } + 9 {B \over k_B} \right) \label{diff7}  
\end{eqnarray}
where $X$ is the number density defined as $X\equiv n_X-n_{\bar X}$; $\partial_\mu X^\mu = v_w {d X \over d \bar z } - D_X {d^2  X \over d \bar z^2 }$, with $D_X$ the diffusion constant of $X$; 
$\Gamma_{ss} = 16 \kappa^\prime \alpha_s^4 T $, being the strong sphaleron rate, with $\kappa^\prime \sim {\cal O } (1)$; 
$\Gamma_{tri}^H$, $\Gamma_{tri}^{S\tilde T}$ and $\Gamma_{tri}^{S \tilde Q}$ are trilinear interaction rates, we refer the reader to Ref.~\cite{Cirigliano:2006wh} for the calculation of these rates in detail.  
We give the values for the statistical factors in the massless limit: $k_{\tilde Q } = 12$, $k_{\tilde T} =6 $, $k_Q= 2 k_T =6$, $k_H=4$ and $k_S=2$.

Solution of the Boltzmann equations can be simplified by the following observations: 
First,  baryon number is locally conserved by neglecting electroweak sphaleron from diffusion equations, $\sum_i (Q_i + U_i+D_i)=0$.
Second, Yukawa interactions of the first two generation quarks can be neglected since they satisfy $1/\Gamma_{Y_{u,d}}\gg \tau_{\rm diff}(\equiv \bar D /v_w^2)$. 
The first two generation quarks can only be generated through the strong sphaleron processes: $\bar t_L t_R \leftrightarrow \bar b_L b_R \sum_i^2 \bar u^i_L u^i_R \bar d^i_L d^i_R$. 
Thus one has $Q_1 =Q_2 = -2 B=2(T+Q)$, where $B$ is the number density of right-handed bottom quark.

Eqs. (\ref{diff1})-(\ref{diff7}) can be solved numerically using the relaxation method, resulting in nonzero $Q$ and $T$. 
The left-handed charge density that biases the weak sphaleron transitions, can be written as $n_L(z)=5Q(z)+4T(z)$, which might be converted into a baryon asymmetry through the weak sphaleron process. 
The final baryon number density takes the form~\cite{Huet:1995sh},
\begin{eqnarray}
n_B=-n_F {\Gamma_{ws} \over 2 v_w } \int_{-\infty}^0 dz n_L (z) \exp\left({15 \Gamma_{ws} \over 4 v_w}z\right)
\end{eqnarray}
which describes baryon generation and washout ahead of the bubble wall.  $\Gamma_{ws} = 6 \kappa \alpha_w^4 T $;  $v_w$ is the bubble wall velocity; $n_F=3$ being the generation of fermions.

\begin{figure}[t]
  \includegraphics[width=0.45\textwidth]{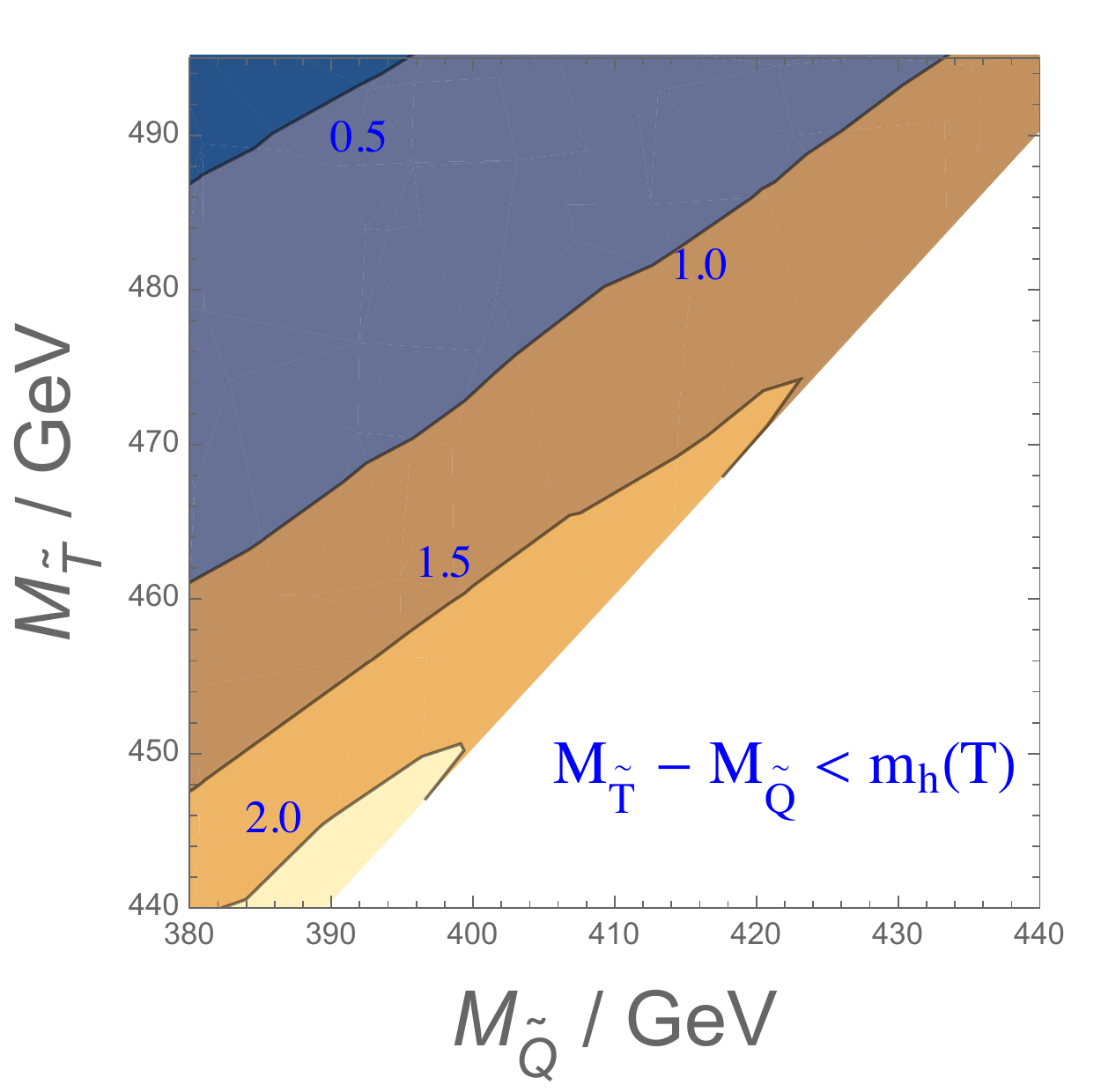}
\caption{\label{ewbg}  Contours of $Y_B^{}/Y_B^{\rm obs}$ in the $M_{\tilde Q} -M_{\tilde T} $ plane by assuming $A_3=120~{\rm GeV}$, $A_4=0.15$ and ${\rm Arg} [A_3^* A_4^{} ]=-{\pi/3}$. For inputs of bubble parameters, see text for detail. }
\end{figure}

We show in Fig.~\ref{ewbg} contours of the scaled baryon asymmetry, $Y_B/Y_B^{\rm obs}$, in the $M_{\tilde Q}-M_{\tilde T} $ plane, by assuming $v_w=0.25$ and $l_w=15/T$, which is the bubble wall width, $v_s(T)=150~{\rm GeV}$, $v_h(T)=100~{\rm GeV}$, $|A_3|=120~{\rm GeV}$, $A_4=0.15$ and ${\rm Arg}[A_3^*A_4^{}]=-\pi/3$.
Contours from the top to the bottom correspond to $Y_B/Y_B^{\rm obs} =0.5,~1.0,~1.5$ and $2.0$ respectively. 
For the uncolored region in the bottom-right corner, one has $M_{\tilde T} -M_{\tilde Q} <m_h^{} (T)$, where $m_h^{} (T) $ is the mass of the SM Higgs at the finite temperature, and thus the trilinear interaction is quenched, which results in a null baryon asymmetry. 
As was shown in Fig.~\ref{signal}, it is possible to generate large enough diphoton signal as $M_X <480~{\rm GeV}$, such that one might have both significant baryon asymmetry and diphton excess in the bottom-left region of the Fig.~\ref{ewbg}.

Finally we study constraints on the model from Higgs measurements, electric dipole moment (EDM) measurements and collider searches of exotic particles.  
The trilinear interaction, $A_3\tilde Q^\dagger H \tilde T$, contributes to both $h\to gg$ and $h\to \gamma \gamma$. It has $\mu_{\gamma \gamma } = 1.17\pm0.27$ for ATLAS~\cite{Aad:2014eha} and $1.14^{+0.26}_{-0.23}$ for the CMS~\cite{Khachatryan:2014ira}. 
Taking $M_{\tilde T}=450~{\rm GeV}$, $M_{\tilde Q} =380~{\rm GeV}$ and $|A_3|=120~{\rm GeV}$, we have $\mu_{\gamma \gamma }=0.998$, which is consistent with the current bound,  and $\mu_{gg} =0.992$, which might be checked by the future Higgs factory CEPC.
In MSSM squark loops may generate an elementary EDM, $d_f^E$, {\it via} two loop Barr-Zee diagram~\cite{Barr:1990vd}, and $d_f^E$ is proportional to  the parity-violating coupling of Higgs with $f$, $g^P_{h\bar f f}$~\cite{Ellis:2008zy}.  
Since there is no $g^P_{h \bar f f}$ in our model, the CP phase from the trilinear interactions receives null constraint from EDMs. 
Besides, scalar quarks should couple to the hidden sector, resulting in quark flavored dark matter, so as to avoid the problem raised by long lived charged particles.   
This interaction is helpful in eluding the collider searches of exotic quarks, similar to the case to stealth supersymmetry~\cite{Fan:2011yu}. 
We leave the collider searches of the model to a future work.

\noindent {\bf {\color{blue}Conclusion}} 
In this letter, we explained both the 750 GeV diphoton excess and BAU in a general NMSSM induced model.
The resonance is a pseudo-scalar singlet, produced through loop-induced gluon fusion with scalar quarks running in the loop. 
The model accommodates a two step EWPT phase transition which can be strongly first order. 
In our model, charge asymmetries are generated via CP-violating interactions of scalar quarks at the expanding bubble wall, which might be transferred into an asymmetry of the Higgs density, then be converted into nonzero number densities of left-handed fermions, resulting BAU via the weak sphaleron process.  
An extensive study of the phenomenology of the model will appear in a forthcoming publication.

\begin{acknowledgments}
This work was supported in part by DOE Grant DE-SC0011095.
\end{acknowledgments}

\end{document}